\documentclass[twocolumn,superscriptaddress,showpacs,preprintnumbers,amsmath,amssymb,aps,pre]{revtex4}
\usepackage{graphicx}% Include figure files
\usepackage{dcolumn}% Align table columns on decimal point
\usepackage{bm}% bold math
\begin{document}
\title{Seismic Electric Signals and 1/f ``noise'' in natural time}
\author{P. A. Varotsos}
\email{pvaro@otenet.gr}
\affiliation{Solid State Section and Solid Earth Physics Institute, Physics Department, University of Athens, Panepistimiopolis, Zografos 157 84,
Athens, Greece}
\author{N. V. Sarlis}
\affiliation{Solid State Section and Solid Earth Physics Institute, Physics Department, University of Athens, Panepistimiopolis, Zografos 157 84,
Athens, Greece}
\author{E. S. Skordas}
\affiliation{Solid State Section and Solid Earth Physics Institute, Physics Department, University of Athens, Panepistimiopolis, Zografos 157 84,
Athens, Greece}

\begin{abstract}
By making use of the concept of natural time, a simple model is
proposed which exhibits the $1/f^a$ behavior with $a$ close to
unity. The properties of the model are compared to those of the
Seismic Electric Signals (SES) activities that have been found to
obey the ubiquitous $1/f^a$ behavior  with $a \approx 1$. This
comparison, which is made by using the most recent SES  data (that
were followed by three magnitude 6.0-class earthquakes), reveals
certain similarities, but the following important difference is
found: The model suggests that the entropy $S_-$ under time
reversal becomes larger compared to the entropy $S$ in forward
time, thus disagreeing with the experimental SES results which
show that $S$ may  be either smaller or larger than $S_-$. This
might be due to the fact that SES activities exhibit {\em
critical} dynamics, while the model cannot capture all the
characteristics of such dynamics.
\end{abstract}

% Keywords required only for MST, PB, PMB, PM, JOA, JOB? %
\maketitle

\section{Introduction}
Among the different features that characterize complex physical
systems, the most ubiquitous is the presence of $1/f^a$ noise in
fluctuating physical variables\cite{MAN99}. This means that the
Fourier power spectrum $S(f)$ of fluctuations scales with
frequency $f$ as $S(f) \sim 1/f^a$. The power-law behavior often
persists over several orders of magnitude with cutoffs present at
both high and low frequencies. Typical values of the exponent $a$
approximately range between 0.8 and 4 (e.g., see Ref.\cite{ANT02}
and references therein), but in a loose terminology   all these
systems are said to exhibit $1/f$ ``noise''. Such a ``noise'' is
found in a large variety of systems, e.g., condensed matter
systems (for example, an excellent review can be found in Ref.\cite{WEI88}), freeway
traffic\cite{MUS76,NAG95,ZHA95}, granular flow\cite{NAK97}, DNA
sequence\cite{gol02}, heartbeat\cite{PEN93}, ionic current
fluctuations in membrane channels\cite{MER99}, river
discharge\cite{MAN69b}, the number of stocks traded
daily\cite{LIL00}, chaotic quantum
systems\cite{GOM05,REL02,SAN05,SAN06}, the light of
quasars\cite{PRE78}, human cognition\cite{GIL95} and
coordination\cite{YOS00}, burst errors in communication
systems\cite{BER63}, electrical measurements\cite{KOG96}, the
electric noise in carbon nanotubes\cite{COL00} and in nanoparticle
films\cite{KIS97}, the occurrence of earthquakes\cite{SOR00} etc.
In some of these systems, the exponent $a$ was reported to be very
close to 1, but good quality data  supporting  such a value exist
in a few of them\cite{WEI88}. As a first example, we refer to the
voltage fluctuations when current flows through a
resistor\cite{YAK00}. As a second example we mention the case of
Seismic Electric Signals (SES) activities which are transient low
frequency ($\leq$ 1Hz) electric signals observed before earthquakes
\cite{proto,var86b,var88x,VAR91,VAR93,var99,grl,JAP2,JAP3}, since
they are emitted when the stress in the focal region reaches a
{\em critical} value before the failure\cite{varbook,newbook}.
These electric signals, for strong earthquakes with magnitude 6.5
or larger, are also accompanied by detectable magnetic field
variations\cite{sar02,PJA1,PJA2,PRL03}.  Actually, the analysis of the
original time series of the SES activities have been shown to obey
a $1/f$-behavior\cite{NAT02,WER05}.

\begin{figure*}
\includegraphics{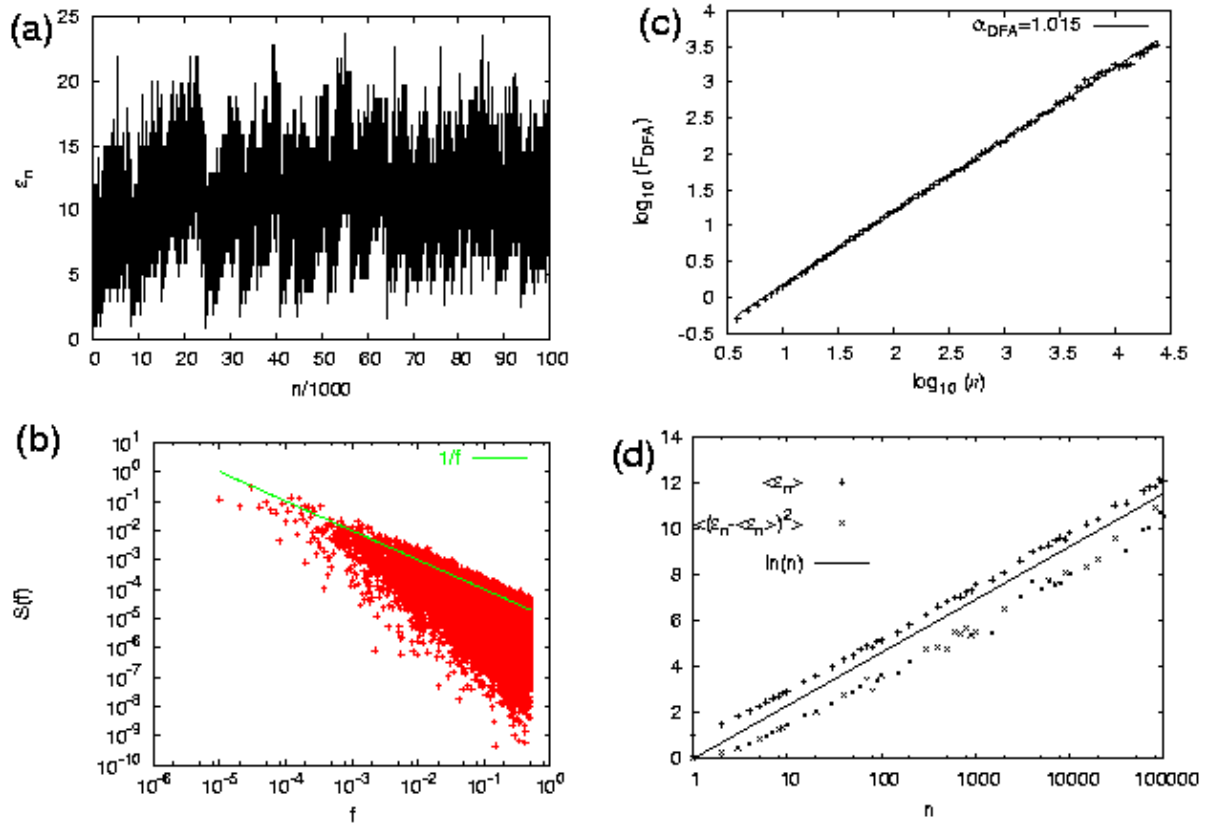}
\caption{(color online) (a):Example of the evolution of
$\epsilon_n$  versus the number of renewals $n$, i.e., in natural
time. An exponential PDF has been considered for the selection of
$\eta_n$ (see the text). (b): The Fourier power spectrum of (a);
the (green) solid line corresponds to $1/f$ and was drawn as a
guide to the eye. (c): The DFA of (a) that exhibits an exponent
$\alpha_{DFA}$ very close to unity, as expected from (b).
(d):Properties of the distribution of $\epsilon_n$. The average
value $\langle \epsilon_n \rangle$ (plus) and the variance
$\langle \left( \epsilon_n -\langle \epsilon_n \rangle \right)^2
\rangle$ (crosses) as a function of n. The straight solid line
depicts $\ln (n)$ and was drawn for the sake of reader's
convenience.} \label{fmodel}
\end{figure*}

%\clearpage

\begin{figure}
 \includegraphics{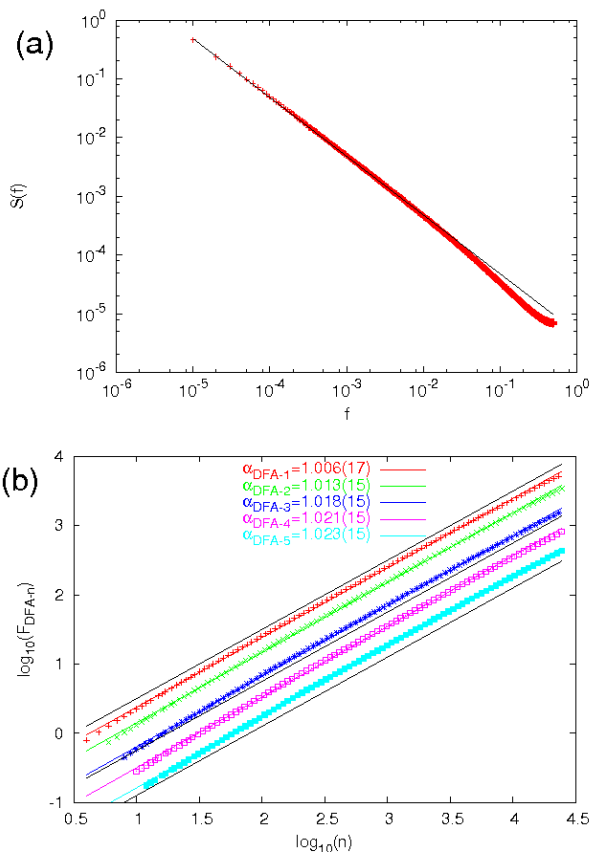}
\caption{(color online) Results from $10^4$ runs of the model
presented in Fig.\ref{fmodel}: (a) the average power spectrum, (b)
Detrended Fluctuation Analyses of order $l$ (DFA-$l$)\cite{Hu01}.
The black solid line in (a) corresponds to $1/f$ spectrum and was
drawn as a guide to the eye. For the same reason in (b), the black
solid lines correspond to $\alpha_{DFA}=1$. In (b), the colored
solid lines correspond to the least square fit of the average
$F_{DFA-l}$, depicted by symbols of the same color; the numbers in
parentheses denote the standard deviation of $\alpha_{DFA-l}$
obtained from the $10^4$ runs of the model. The various
$F_{DFA-l}$ have been displaced vertically for the sake of
clarity.}
 \label{sfdfa}
\end{figure}

\begin{figure}
 \includegraphics{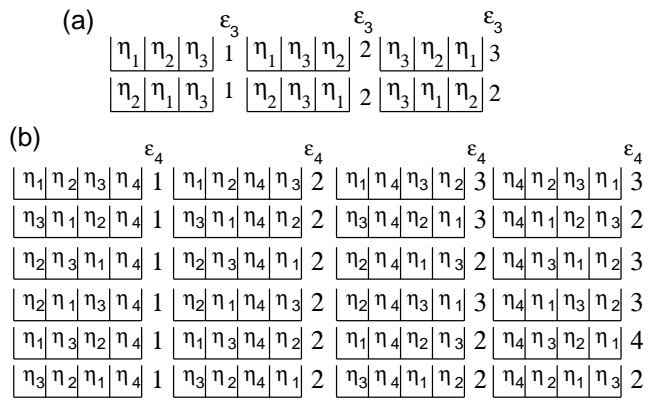}
\caption{The $\eta_n$ values arranged in sites (bins) according to
their value increasing from left to right. (a)The six(=3!) equally
probable outcomes after the selection of 3 random numbers by the
same PDF. Actually, the sample space is (in one to one
correspondence to) the permutations of 3 objects. (b) The 24(=4!)
equally probable outcomes after the selection of 4 random numbers
by the same PDF. Again, the sample space is (in one to one
correspondence to) the permutations of 4 objects. For the reader's
convenience, in each outcome, the corresponding $\epsilon_n$-value
($n=3$ or 4) is written. An inspection of (b), shows that
$p(\epsilon_4=1)=1/4, p(\epsilon_4=2)=11/24,  p(\epsilon_4=3)=1/4$
and $p(\epsilon_4=4)=1/24$. }
 \label{x4}
\end{figure}

\begin{figure}
 \includegraphics{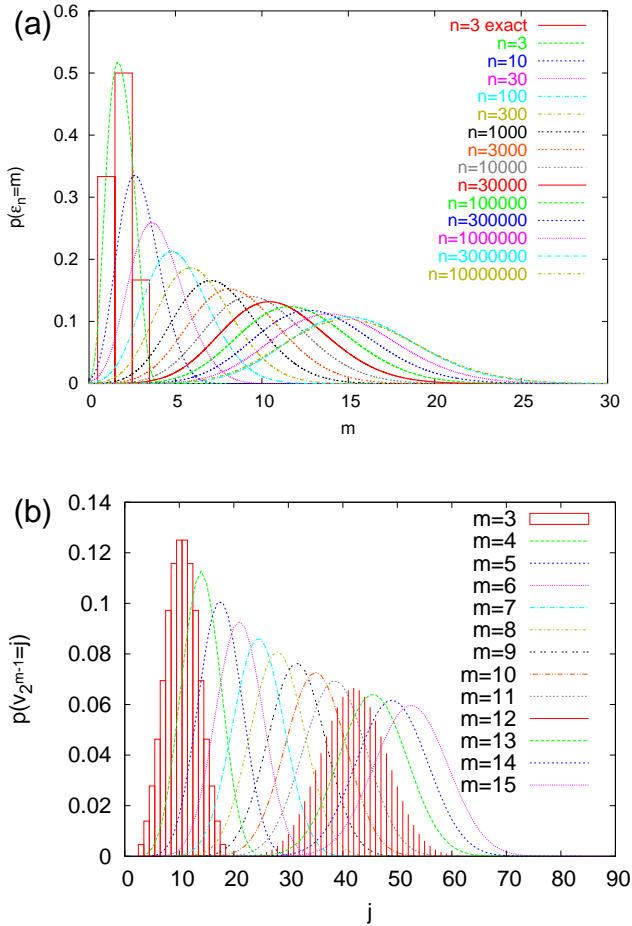}
\caption{(color online) (a)The probabilities $p(\epsilon_n=m)$ as
a function of $m$ for various $n$. The bar chart corresponds to
the exact $p(\epsilon_3=m)$ whereas the continuous lines to the
Cornish-Fisher approximation of Eq.(\ref{appp}). The latter
approximation converges very rapidly to the true
$p(\epsilon_n=m)$, see for example $n=3$. This fact enables the
calculation of $p(\epsilon_n=m)$ for very large $n$, for which the
recursive relation of Eq.(\ref{dom}) would accumulate significant
round-off errors. (b) The probabilities $p(v_{2^{m-1}}=j)$ as a
function of $j$ for various $m$ (see Appendix B). They are clearly
skewnessless, i.e., symmetric around their mean. Here, as in
Ref.\cite{voss}, 6-sided dices ($k=6$) were considered. For the
reader's convenience $p(v_{2^{m-1}}=j)$ for $m=3$ and $m=12$ have
been drawn with a different style.}
 \label{pdf}
\end{figure}

\begin{figure}
\includegraphics{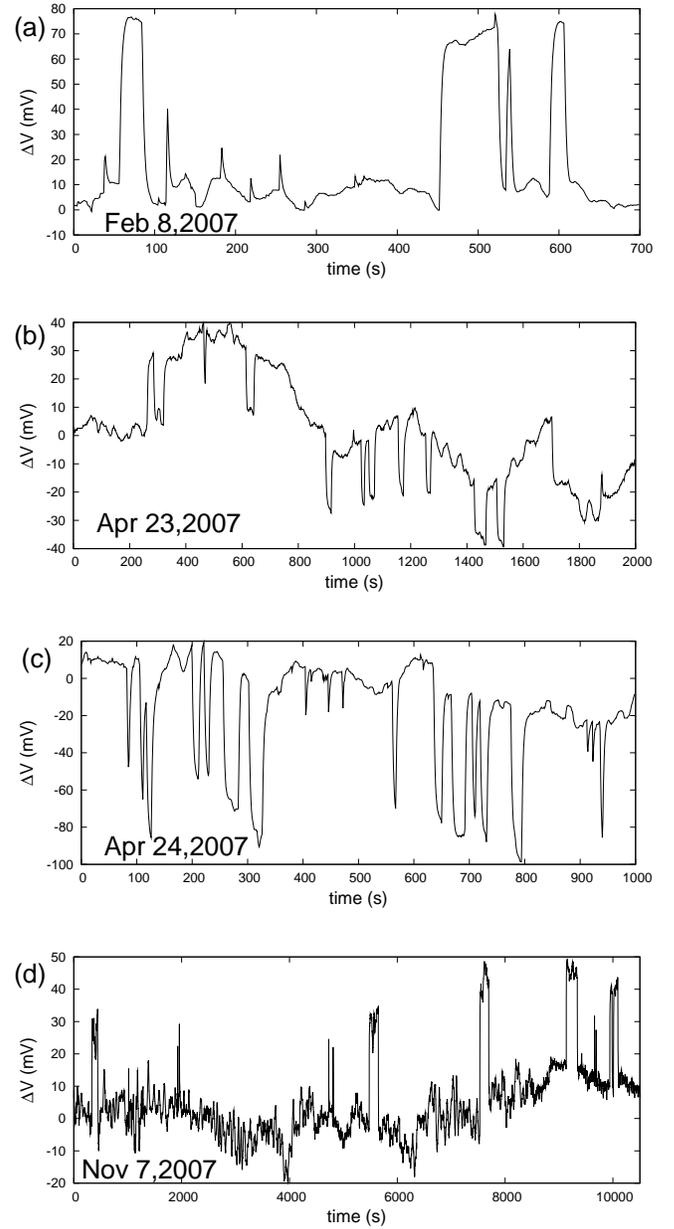}
\caption{Four electric signals recorded at PAT(sampling rate
$f_{exp}$=1 sample/sec) on February 8, 2007(a), April 23, 2007(b),
April 24, 2007(c) and November 7, 2007(d). } \label{f3}
\end{figure}

\begin{figure}
\includegraphics{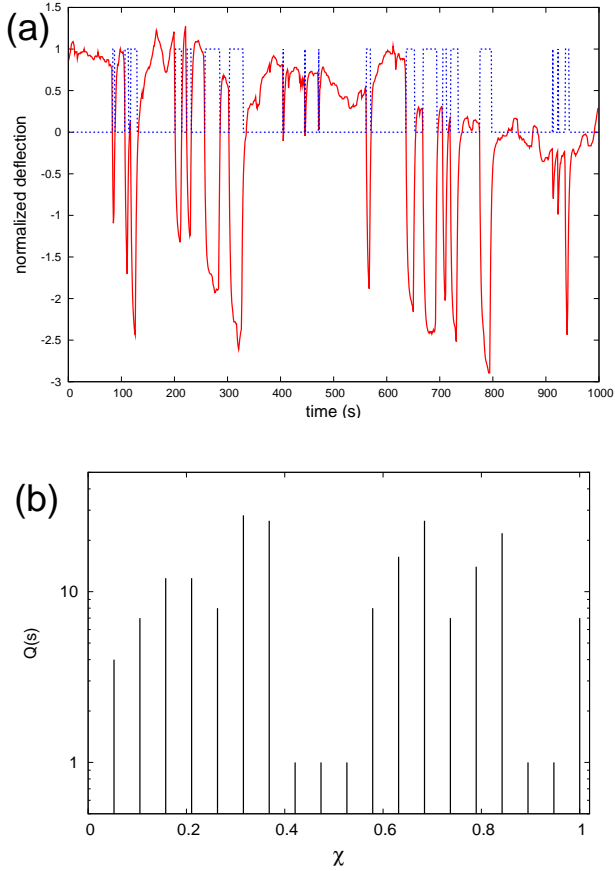}
\caption{(color online) (a): The electric signal depicted in
Fig.\ref{f3}(c) (April 24,2007) in normalized units  (i.e., by
subtracting the mean value and dividing the results by the
standard deviation) along with its dichotomous representation
which is marked by the dotted (blue) line. (b): How the signal in
(a) is read in natural time. } \label{fn6}
\end{figure}

\begin{figure}
 \includegraphics{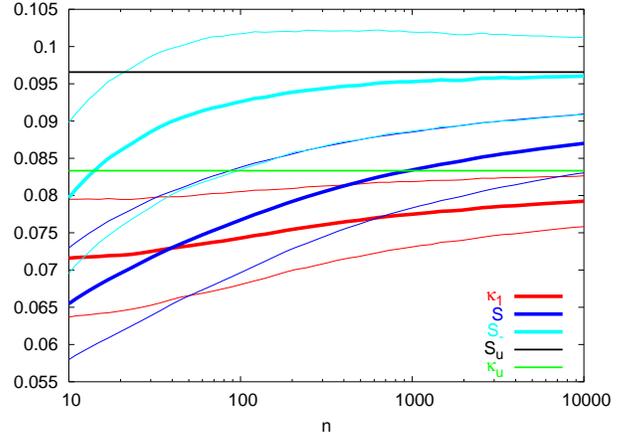}
\caption{(color online) Evolution of the parameters of $\kappa_1$,
$S$ and $S_-$  as a function of $n$, when $\epsilon_n$ are
analyzed in the natural time domain.  The thick lines correspond
to the average value of $\kappa_1$, $S$ and $S_-$, found by $10^4$
runs of the model. The thinner lines correspond to the $\pm$one
standard deviation confidence intervals. For the reader's
convenience, the green and black horizontal lines show the values
$\kappa_u$ and $S_u$ of $\kappa_1$ and $S$, respectively, that
correspond to a ``uniform''  distribution. }
 \label{natmodel}
\end{figure}

\begin{figure}
\includegraphics{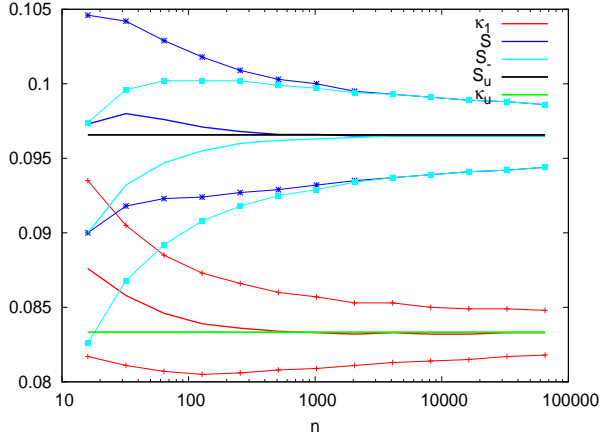}
\caption{(color online) Evolution of the parameters of $\kappa_1$,
$S$ and $S_-$  as a function of $n$, when $v_n$  (see Appendix B)
are analyzed in the natural time domain.  Since the model produces
1/f noise time series with lengths $n=2^m$, the calculation was
performed at such $n$ (i.e., $n=10^m$, $m=4,5, \ldots 16$)
indicated by the  points marked. The thick lines connect the
corresponding points of the average value of $\kappa_1$, $S$ and
$S_-$, found by $10^4$ runs of the model. The thinner lines
connect the  points corresponding to the $\pm$one standard
deviation confidence intervals. For the reader's convenience, the
green and black horizontal lines show the values $\kappa_u$ and
$S_u$ of $\kappa_1$ and $S$, respectively, that correspond to a
``uniform'' distribution. Here, as in Ref.\cite{voss}, 6-sided
dices ($k=6$) were considered. } \label{voss}
\end{figure}

\begin{figure}
\includegraphics{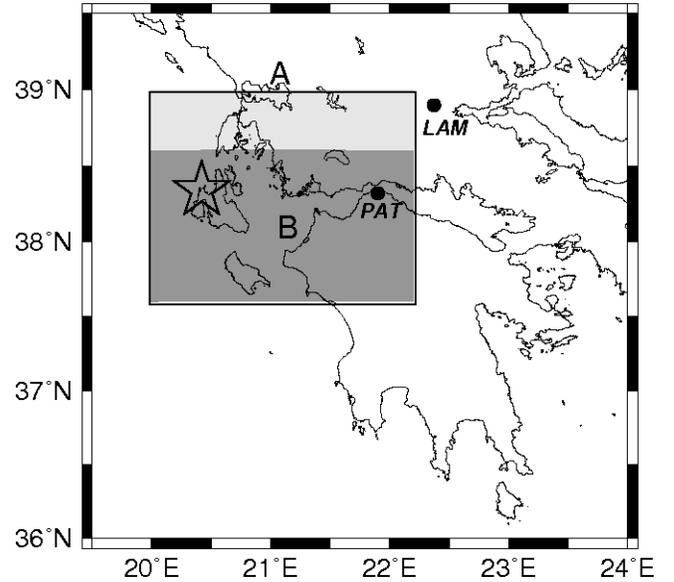}
\caption{The map shows the areas A,B. The star indicates the
epicenter of the strong 6.0 EQ that occurred on March 25, 2007 in
Kefallonia.} \label{af1}
\end{figure}

 \begin{figure*}
\includegraphics{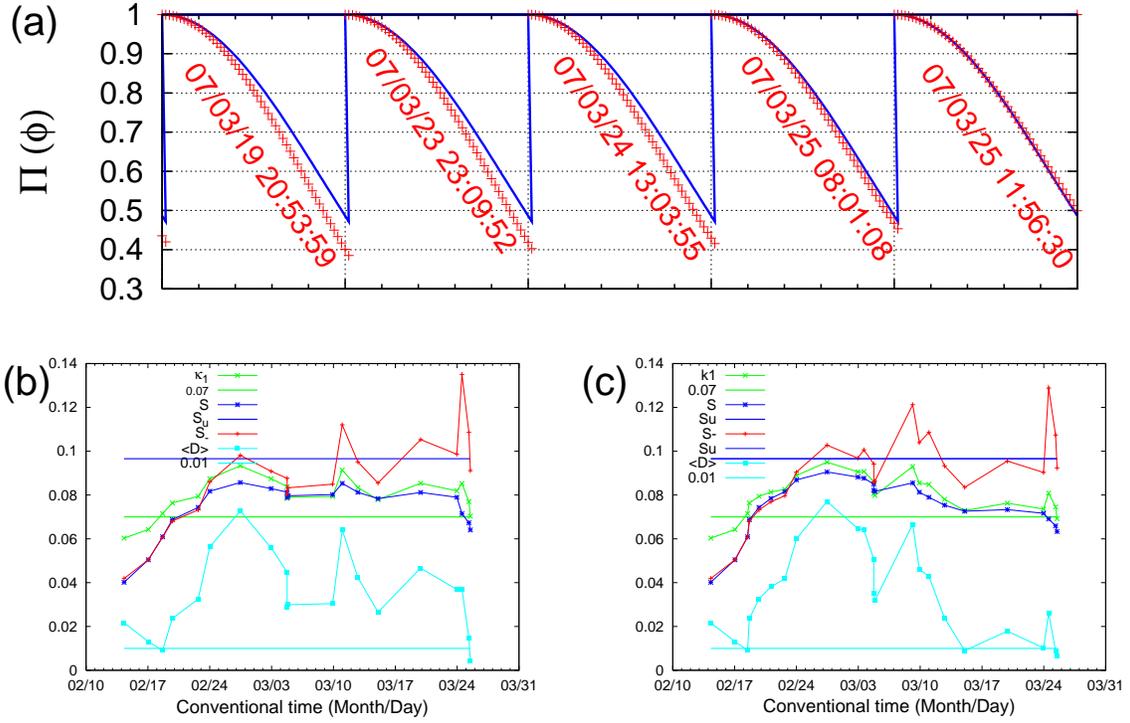}
\caption{(color online) (a) The normalized power spectrum(red)
$\Pi (\phi )$ of the seismicity   as it evolves event by event
(whose date and time (UT) of occurrence are written in each panel)
after the initiation of the SES activity on February 8, 2007.  The
excerpt presented here refers to the period 19 to 25 March, 2007
and corresponds to the area B, $M_{thres}=3.2$.  In each case only
the spectrum in the area $\phi \in [0,0.5]$ is depicted (separated
by the vertical dotted lines), whereas the  $\Pi (\phi )$ of
Eq.(2) is depicted by blue color. The minor horizontal ticks for
$\phi$ are marked every 0.1. (b), (c) Evolution of the parameters
$\langle D \rangle$, $\kappa_1$, $S$ and $S_{-}$ after the
initiation of the SES activity on February 8, 2007 for the areas B
($M_{thres}=3.2$) and A($M_{thres}=3.2$), respectively, until just
before the 6.0 EQ.} \label{af2}
\end{figure*}

 \begin{figure}
\includegraphics{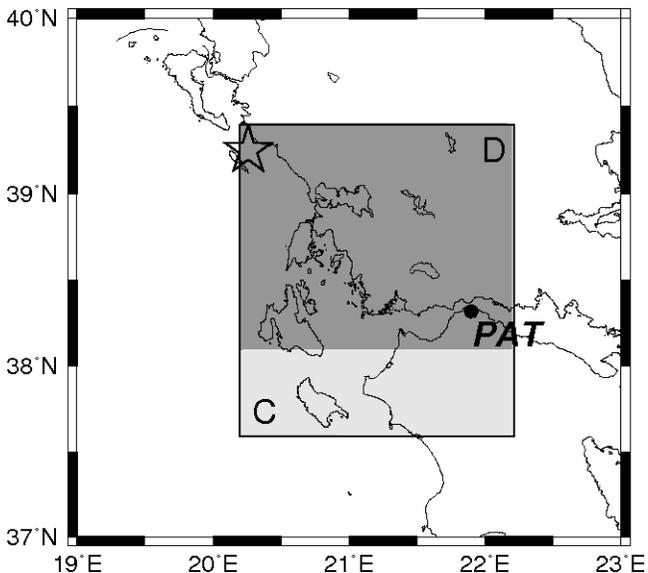}
\caption{The map shows the areas C and D. The star shows the
epicenter of the strong 5.8 EQ at 18:09:11 on June 29, 2007.}
\label{afx1}
\end{figure}

 \begin{figure*}
\includegraphics{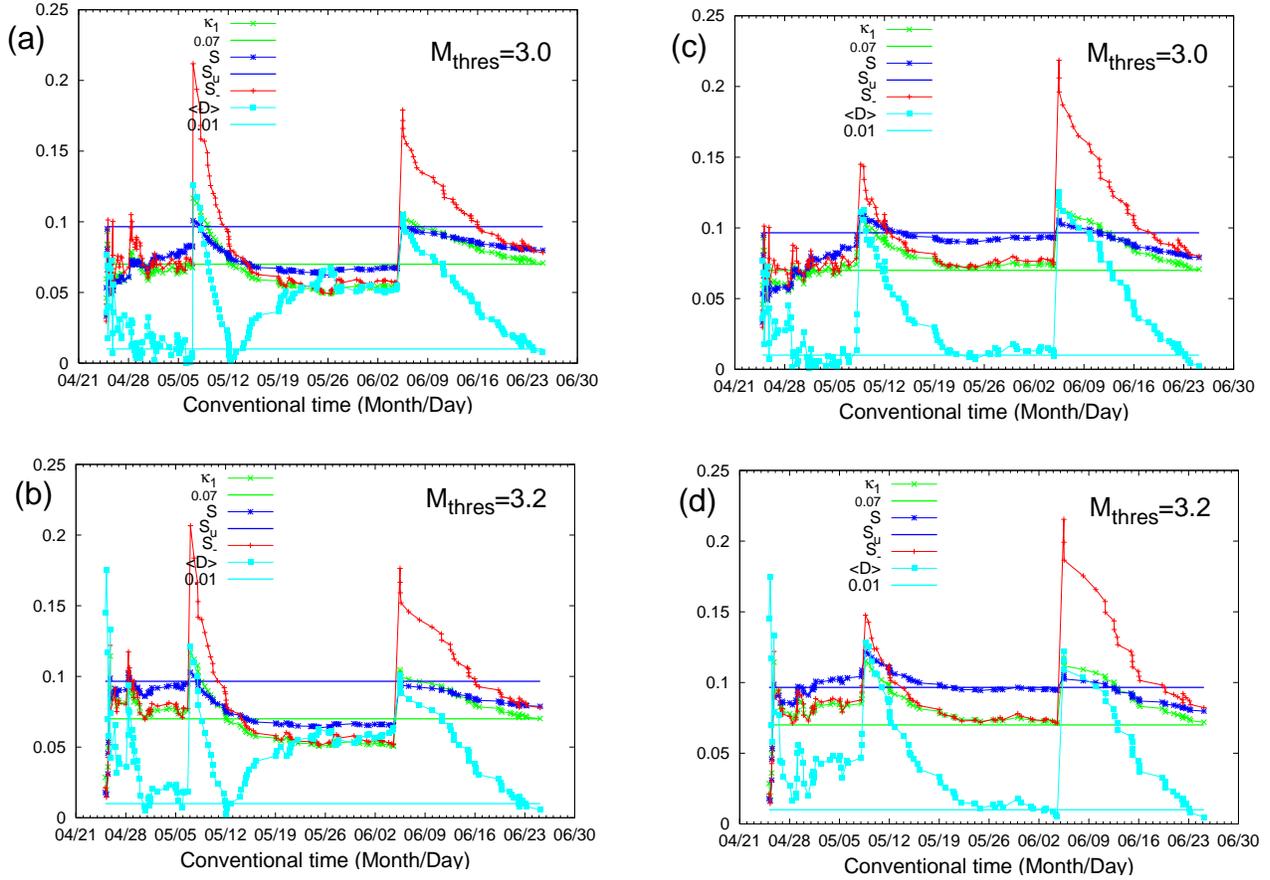}
\caption{(color online) (a),(b) and (c),(d) depict the evolution
of the parameters  $\langle D \rangle$, $\kappa_1$, $S$ and
$S_{-}$ after the initiation of the SES activity on April 24, 2007
for the areas C  and D, respectively (for two magnitude thresholds
in each area),   until 03:40:15 UT on June 25, 2007.} \label{afx2}
\end{figure*}

 \begin{figure}
\includegraphics{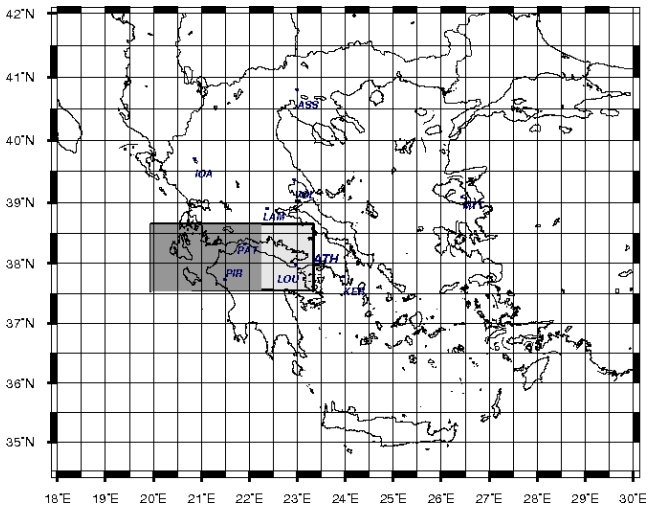}
\caption{(color online) The map shows the areas
$N^{38.6}_{37.6}E^{22.2}_{20.0}$ and
$N^{38.6}_{37.6}E^{23.3}_{20.0}$ in which the seismicity was
studied\cite{var07x} after the SES activity recorded at PAT on 7
November, 2007. The solid dots stand for the sites at which
electric field variations are continuously monitored with a
sampling frequency 1Hz.} \label{fig12}
\end{figure}

\begin{figure}
\includegraphics{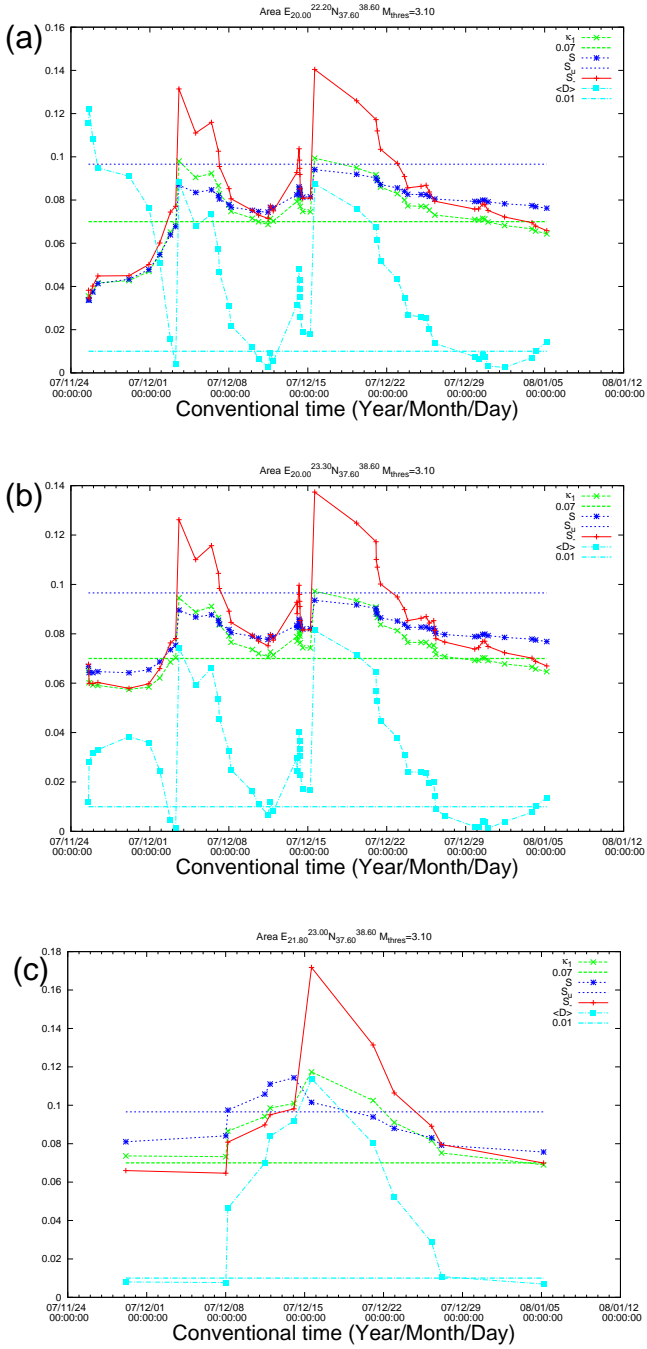}
\caption{(color online) The evolution of the parameters $\langle D
\rangle$, $\kappa_1$, $S$ and $S_{-}$ for the period: since the
initial submission of this paper until just before the 6.6
earthquake on 6 January, 2008. (a), (b) correspond to the two
areas depicted in Fig.\ref{fig12}, while (c) to the dark shaded
area in Fig.\ref{fig14}, $M_{thres}=3.1$.  } \label{fig13}
\end{figure}

\begin{figure}
\includegraphics{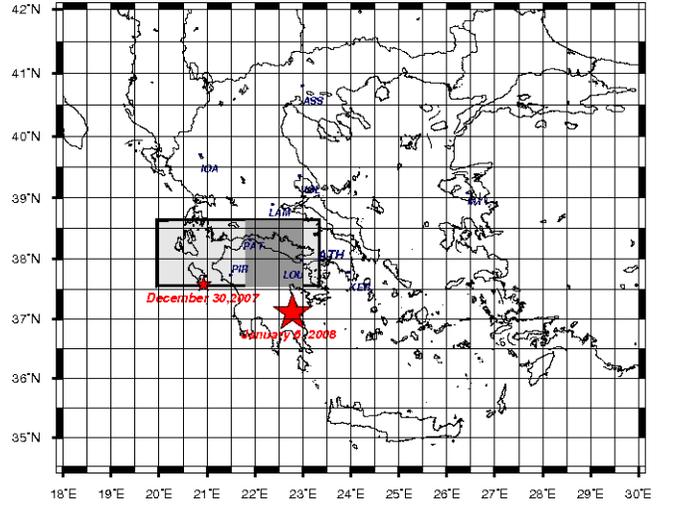}
\caption{(color online) The map shows the epicenters of the 5.3
earthquake on 30 December, 2007 (small star) and the 6.6
earthquake on 6 January, 2008 (large star). The dark shaded area
depicts the part of the larger area
$N^{38.6}_{37.6}E^{23.3}_{20.0}$ adjacent to the epicenter of the
6.6 earthquake, whose the four parameters $\langle D \rangle$,
$\kappa_1$, $S$ and $S_{-}$ fullfilled the conditions for a true
coincidence just one day before the major earthquake occurrence
(see Fig.\ref{fig13}(c)).}\label{fig14}
\end{figure}

\begin{figure}
\includegraphics{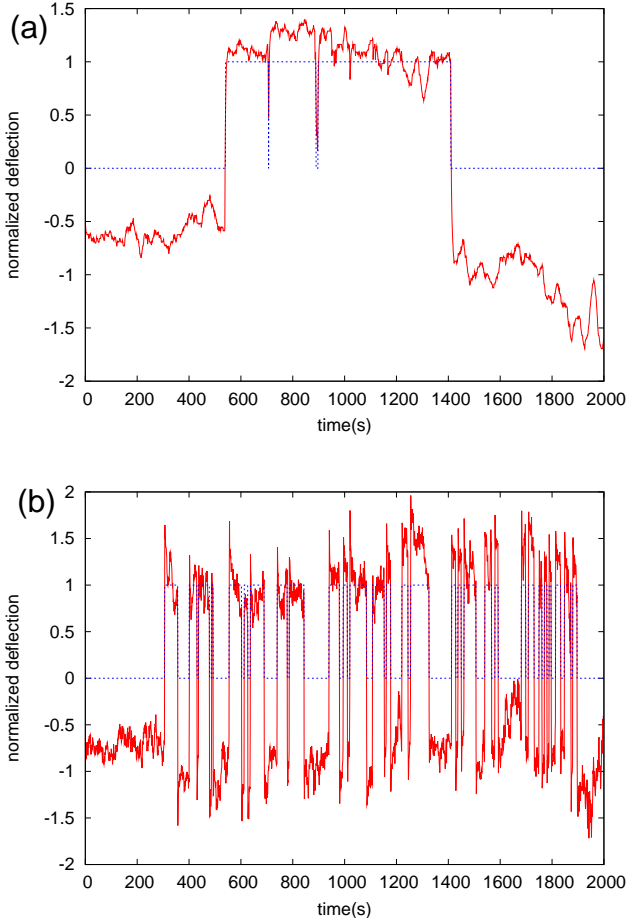}
\caption{(color online) The electric signals recorded on 10
January, 2008 at PAT (a) and on 14 January, 2008 at PIR (b), in a
fashion similar to that of Fig.\ref{fn6}(a) }\label{NFX}
\end{figure}

\begin{figure}
\includegraphics{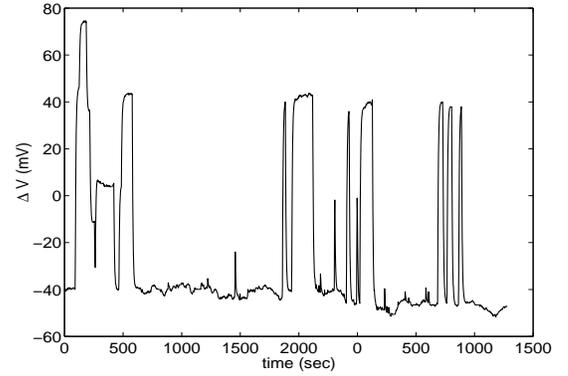}
\caption{The electric signal recorded on 9 October, 2008 at PAT in
a fashion similar to that of Fig.\ref{f3}}\label{R091008}
\end{figure}

\begin{table}
\caption{ The values of $S$, $\kappa_1$, $S_-$ for the electric
signals presented in Fig.\ref{f3}.} \label{tab1}
\begin{ruledtabular}
\begin{tabular}{cccc}
%  \hline

 Date recorded & $S$ & $\kappa_1$ & $S_-$  \\
\hline
Feb 8, 2007 &  0.067$\pm$0.007        &   0.074$\pm$0.007  & 0.079$\pm$0.007  \\
Apr 23, 2007  & 0.071$\pm$0.005        &  0.069$\pm$0.003 & 0.066$\pm$0.005 \\
Apr 24, 2007 &   0.072$\pm$0.003     &  0.067$\pm$0.003 & 0.069$\pm$0.003 \\
Nov 7, 2007  & 0.070$\pm$0.005       &  0.065$\pm$0.005 &  0.070$\pm$0.005\\
\end{tabular}
\end{ruledtabular}
\end{table}

The $1/f^a$ behavior has been well understood on the basis of
dynamic scaling observed at {\em equilibrium} critical points
where the power-law correlations in time stem from the
infinite-range correlations in space (see Ref.\cite{ANT02} and
references therein). Most of the observations mentioned above,
however, refer to {\em nonequilibrium} phenomena for which
-despite some challenging theoretical
attempts\cite{BAK87,BAK96,ANT01,DAV02}- possible {\em generic}
mechanisms leading to scale invariant fluctuations have not yet
been identified. In other words, despite its ubiquity, there is no
yet universal explanation about the phenomenon of the $1/f^a$
behavior. Opinions have been expressed (e.g., see
Ref.\cite{GOM05}) that it does not arise as a consequence of
particular physical interactions, but it is a generic
manifestation of complex systems.

It has been recently
shown\cite{NAT01,NAT02,NAT02A,NAT03,NAT03B,NAT04,NAT05,NAT05B,VAR05C,VAR06PRB,NAT06A,NAT06B,APLHeart}
that novel dynamic features hidden behind the time series of
complex systems can emerge if we analyze them in terms of a newly
introduced time domain, termed natural time $\chi$ (see below). It
seems that this analysis enables the study of the dynamic
evolution of a complex system and identifies when the system
enters a critical stage.  Natural time domain is
optimal\cite{ABE05} for enhancing the signal's localization in the
time frequency space, which  conforms to the desire to reduce
uncertainty and extract signal information as much as possible. In
a time series comprising $N$ events, the {\em natural time}
$\chi_k = k/N$ serves as an index\cite{NAT01,NAT02,NAT02A} for the
occurrence of the $k$-th event. The evolution of the pair
($\chi_k, Q_k$) is
studied\cite{NAT01,NAT02,NAT02A,NAT03,NAT03B,NAT04,NAT05,NAT05B,VAR05C,NAT06A,NAT06B,newbook},
where $Q_k$ denotes a quantity proportional to the {\em energy}
released in the $k$-th event.

 The scope of the present paper is
twofold. First,  a simple model is proposed (Section \ref{dyo})
which, in the frame of natural time, leads to $1/f^a$ behavior
with an exponent $a$ close to unity. Second, the properties of this model
in natural time are
 compared to those of the SES activities in Section \ref{tria}. This comparison is carried out
by making use of the  most recent experimental data of SES
activities observed in Greece during the last several months.
Section \ref{tesera} presents the conclusions. Two Appendices are
also provided, the first of which refers to the earthquakes that
followed the SES activities presented here. Appendix B clarifies
that an early model proposed by Voss\cite{voss} differs
essentially from the one presented here.

\section{The model proposed}
\label{dyo} \subsection{Description of the model} Here, we present
a  simple competitive evolution model which results, when analyzed
in natural time, to $1/f^a$ ``noise'' with $a$ very close to
unity. Let us consider the cardinality $\epsilon_n$ of the family
of sets ${\rm E_n}$ of {\em successive extrema} obtained from a
given probability distribution function (PDF); ${\rm E_0}$ equals
to the empty set. Each ${\rm E_n}$ is obtained by following the
procedure described below for $n$ times. Select a random number
$\eta_n$ from a given PDF (here, we use the exponential PDF, i.e.,
$p(\eta_n)=\exp(-\eta_n)$) and compare it with all the members of
${\rm E_{n-1}}$. In order to construct the set ${\rm E_{n}}$, we
{\em disregard} from the set ${\rm E_{n-1}}$ all its members that
are smaller than $\eta_n$ and furthermore {\em include} $\eta_n$.
Thus, ${\rm E_{n}}\neq \varnothing$ for {\em all} n$>0$ and ${\rm
E_{n}}$ is a finite set of real numbers whose members are always
larger or equal to $\eta_n$. Moreover, $\min[ {\rm E_n}] \geq
\min[{\rm E_{n-1}}]$ and $\max [{\rm E_n}] \geq \max[{\rm
E_{n-1}}]$. The cardinality $\epsilon_n \equiv \left| {\rm  E_n
}\right|$ of these sets, which may be considered as equivalent to
the dimensionality of the thresholds distribution in the coherent
noise model (e.g. see Ref.\cite{ABE04} and references therein), if
considered as time-series with respect to the {\em natural} number
$n$ (see Fig.\ref{fmodel}(a)) exhibits $1/f^a$ noise with $a$ very
close to unity, see Fig.\ref{fmodel}(b). This very simple model
whose evolution is depicted in Fig.\ref{fmodel}(a), leads to a
Detrended Fluctuation Analysis\cite{PEN93} (DFA) exponent
$\alpha_{DFA}$ close to unity, see Fig.\ref{fmodel}(c), being
compatible with the $1/f$ power spectrum depicted in
Fig.\ref{fmodel}(b). The mathematical model described above
corresponds to an asymptotically non-stationary process, since
$\langle \epsilon_n \rangle \propto \ln n$ with a variance
$\langle \left( \epsilon_n -\langle \epsilon_n \rangle \right)^2
\rangle \propto \ln n$ (see Fig.\ref{fmodel}(d)). Thus, in simple
words, the present  model suggests that the cardinality
$\epsilon_n$ of the family of sets ${\rm E_n}$ of successive
extrema exhibits $1/f^a$ behavior when considered as time-series
with respect to the natural (time) number $n$. We note that a
connection between $1/f^a$ noise and extreme value statistics has
been established and proposed as providing a new angle at the
generic aspect of the phenomena\cite{ANT01}.  Furthermore, in the
frame of a formal similarity between the discrete spectrum of
quantum systems and a discrete time series\cite{REL02} the
following striking similarity is noticed:  The fact that $a\approx
1$ together with the behavior $\langle \left( \epsilon_n -\langle
\epsilon_n \rangle \right)^2 \rangle \propto \ln n$ of the present
model  is  reminiscent of the power law exponent and the $\langle
\delta_n^2 \rangle$ statistic in chaotic quantum
systems\cite{REL02,SAN05}.

In order to check the stability of the results of
Fig.\ref{fmodel},  we present in Fig.\ref{sfdfa}(a) the average
power spectrum obtained from $10^4$ runs of the model. A sharp
$1/f$ behavior is observed. Moreover, in Fig.\ref{sfdfa}(b), we
present the results of the corresponding average values of
$F_{DFA-l}$ obtained from DFA of various orders $l$ (i.e., when
detrending with a polynomial of order $l$, see Ref.\cite{Hu01}).
Figure \ref{sfdfa}(b) indicates that $\alpha_{DFA-l}$ is close to
unity.

\subsection{Analytical properties}
\label{dyoB}
We now discuss an analytical procedure which clarifies some
properties of the model. In order to  find  analytically the
distribution of the probabilities $p(\epsilon_n)$, one has simply
to consider the possible outcomes when drawing $n$ random numbers
$\eta_n$. Since the selection is made by a means of a PDF, all
these numbers are different from each other, thus -when sorted
they- are equivalent to  $n$ points (sites) lying on the real axis.
The value of $\epsilon_n$ varies as $\left\{ \eta_n \right\}$
permutate along these  $n$ sites {\em independently} from the PDF
used in the calculation. Thus, a detailed study of the permutation
group of $n$ objects can lead to an exact solution of the model. It
is well known, however, that the number of the elements of this
group is n! and this explains why we preferred to use the
numerical calculation shown in Fig.\ref{fmodel}.  Some exact
results obtained by this method are the following: $\langle
\epsilon_1 \rangle=1$; $\langle \epsilon_2 \rangle=1+1/2$, since
$p(\epsilon_2=1)=p(\epsilon_2=2)=1/2$; $\langle \epsilon_3
\rangle=1+1/2+1/3$, since $p(\epsilon_3=1)=1/3,
p(\epsilon_3=2)=1/2$ and $p(\epsilon_3=3)=1/6$; $\langle
\epsilon_4 \rangle=1+1/2+1/3+1/4$ (see Fig.\ref{x4}). Figure
\ref{x4} analyzes the results for $n=3$ (Fig.\ref{x4}(a)) and
$n=4$ (Fig.\ref{x4}(b)).  One can see that the probability
$p(\epsilon_n=m)$ equals to the sum of the $n$ possible outcomes
as $\eta_n$ moves from the left to right in the $n$ columns of
Fig.\ref{x4}. In each column, the probability to have at the end
$\epsilon_n=m$ is just equal to the probability to keep  $m-1$
numbers from the numbers already drawn that are larger than
$\eta_n$. This results in
\begin{equation}
p(\epsilon_n=m)=\frac{1}{n} \sum_{k=m-1}^{n-1} p(\epsilon_k=m-1)
\label{dom}
\end{equation}
(cf. $p(\epsilon_0=0)=1$).

%Using Eq.(\ref{dom}), one can prove that $\langle \epsilon_n
%\rangle =\langle \epsilon_{n-1} \rangle+1/n$, which reflects that
%$\langle \epsilon_n \rangle =\sum_{k=1}^n1/k$.
Equation (\ref{dom}) enables us to calculate the characteristic
function (see p.928 of Ref.\cite{abr70})
\begin{equation}
f_n(\lambda)\equiv \langle \exp \left( \lambda \epsilon_n \right) \rangle=\sum_{m=1}^n e^{\lambda m}p(\epsilon_{n}=m) .
\end{equation}
Indeed,  by substituting $p(\epsilon_{n}=m)$ in $f_{n}(\lambda)$, we obtain
\begin{equation}
nf_n(\lambda)=\sum_{m=1}^n  e^{\lambda m} \sum_{k=m-1}^{n-1} p(\epsilon_k=m-1),
\label{m2}
\end{equation}
whereas by substituting $p(\epsilon_{n+1}=m)$ in $f_{n+1}(\lambda)$, we find
\begin{equation}
(n+1)f_{n+1}(\lambda)= \sum_{m=1}^n e^{\lambda m} \sum_{k=m-1}^{n-1} p(\epsilon_k=m-1)+e^\lambda f_\lambda(n).
\label{m1}
\end{equation}
Subtracting now Eq.(\ref{m2}) from Eq.(\ref{m1}), we finally get
\begin{equation}
f_{n+1}(\lambda)= \frac{n+e^\lambda}{n+1} f_{n}(\lambda).
\label{ana1}
\end{equation}
Since $f_1(\lambda)=e^{\lambda}$, we find that Eq.(\ref{ana1}) -upon considering Eq.6.1.22 of Ref.\cite{abr70}-  results in
\begin{equation}
f_n(\lambda)=\frac{1}{n!}\frac{\Gamma(e^\lambda+n)}{\Gamma(e^\lambda)},
\label{chara}
\end{equation}
where $\Gamma(x)$ is the gamma function. Now, the mean and all the
central moments $\mu_l \equiv\langle \left( \epsilon_n -\langle
\epsilon_n \rangle \right)^l\rangle$ of the distribution of
$p(\epsilon_{n}=m)$ can be obtained by virtue of the cumulant
theorem (see p.928 of Ref.\cite{abr70}):
\begin{eqnarray}
\langle \epsilon_n \rangle= \left. \frac{d }{d \lambda} \ln f_n(\lambda)  \right|_{\lambda=0}, \label{c1} \\
\mu_2\equiv\langle \left( \epsilon_n -\langle \epsilon_n \rangle
\right)^2 \rangle=\left. \frac{d^2}{d \lambda^2}  \ln f_n(\lambda) \right|_{\lambda=0}, \label{c2} \\
\mu_3\equiv\langle \left( \epsilon_n -\langle \epsilon_n \rangle
\right)^3\rangle=\left. \frac{d^3}{d \lambda^3}  \ln f_n(\lambda) \right|_{\lambda=0}, \label{c3} \\
\mu_4-3\mu_2^2=\left. \frac{d^4}{d \lambda^4}  \ln f_n(\lambda) \right|_{\lambda=0}. \label{c4}
\end{eqnarray}
Substituting Eq.(\ref{chara}) into Eqs.(\ref{c1}) to (\ref{c4}) and using the properties of the polygamma functions (i.e., the n-th order  logarithmic derivatives of the gamma function, see p.260 of Ref.\cite{abr70}), we obtain
\begin{eqnarray}
\langle \epsilon_n \rangle= \sum_{k=1}^n \frac{1}{k}, \label{t1} \\
\langle \left( \epsilon_n -\langle \epsilon_n \rangle
\right)^2 \rangle=\sum_{k=1}^n \left( \frac{1}{k} - \frac{1}{k^2} \right), \label{t2} \\
\mu_3=\sum_{k=1}^n \left( \frac{1}{k} - \frac{3}{k^2} + \frac{2}{k^3} \right), \label{t3} \\
\mu_4-3\mu_2^2=\sum_{k=1}^n \left( \frac{1}{k} - \frac{7}{k^2} + \frac{12}{k^3} - \frac{6}{k^4}\right). \label{t4}
\end{eqnarray}
Equations (\ref{t1}) to (\ref{t4}) enable us to calculate the mean,
standard deviation $\sigma$($=\sqrt{\mu_2}$), skewness $\gamma_1=\mu_3/\sigma^3$ and kurtosis $\gamma_2=\mu_4/\sigma^4-3$ as a function of $n$. Using now the Cornish-Fisher (CF) expansion treated in Ref.\cite{kota}, we obtain the following continuous approximation to $p(\epsilon_{n}=m)$
\begin{eqnarray}
p_{CF}(\tilde{\epsilon_{n}})=\frac{1}{\sqrt{2\pi}} \left|
1-\frac{\gamma_1}{3}\tilde{\epsilon_{n}}+\frac{\gamma_1^2}{36} (12 \tilde{\epsilon_{n}}^2-7)-\frac{\gamma_2}{8} (\tilde{\epsilon_{n}}^2-1)\right| \times \nonumber \\
 \times
\exp \left\{ -\frac{1}{2} \left[ \tilde{\epsilon_{n}}-\frac{\gamma_1}{6} (\tilde{\epsilon_{n}}^2-1)-\frac{\gamma_2}{24} (\tilde{\epsilon_{n}}^3-3\tilde{\epsilon_{n}}) +\frac{\gamma_1^2}{36} (4\tilde{\epsilon_{n}}^3-7\tilde{\epsilon_{n}}) \right]^2 \right\},
\label{appp}
\end{eqnarray}
where $\tilde{\epsilon_{n}}=(\epsilon_n-\langle \epsilon_n
\rangle)/\sigma$. Equation (\ref{appp}), although being a
continuous approximation to the point probabilities
$p(\epsilon_{n}=m)$, rapidly converges to the latter, see for
example the comparison of the exact  $p(\epsilon_{3}=m)$  and the
corresponding $p_{CF}(\tilde{\epsilon_{3}})/\sigma$ in
Fig.\ref{pdf}(a). An inspection of this figure, which  depicts the
probabilities $p(\epsilon_{n}=m)$ up to $n=10^7$, reveals that
even for large $n$, the probability $p(\epsilon_{n}=m)$ remains
non-Gaussian (cf. even  at $n=10^9$, we obtain from Eqs.(\ref{t2})
to (\ref{t4})  $\gamma_1=0.2154\neq 0$  with $\gamma_2=0.0459 >
0$).

\section{Comparison of the model with the SES physical properties in natural time}
\label{tria}

 For dichotomous
signals, which is frequently the case of SES activities, the
quantity $Q_k$ mentioned in Section I stands for the duration of
the $k$-th pulse.
 The normalized power spectrum $\Pi(\omega )\equiv | \Phi (\omega ) |^2 $ was
introduced\cite{NAT01,NAT02,NAT02A}, where
\begin{equation}
\label{eq3} \Phi (\omega)=\sum_{k=1}^{N} p_k \exp \left( i \omega
\frac{k}{N} \right)
\end{equation}
and $p_k=Q_{k}/\sum_{n=1}^{N}Q_{n}$, $\omega =2 \pi \phi$; $\phi$
stands for the {\it natural frequency}. The continuous function
$\Phi (\omega )$ should {\em not} be confused with the usual
discrete Fourier transform, which considers only its values at
$\phi=0,1,2,\ldots$. In natural time
analysis\cite{NAT01,NAT02,NAT02A,newbook}, the properties of
$\Pi(\omega)$ or $\Pi(\phi)$ are studied  for natural frequencies
$\phi$
 less than 0.5, since in
this range of $\phi$, $\Pi(\omega)$  or $\Pi(\phi)$ reduces
 to a {\em characteristic function} for the
probability distribution $p_k$  in the context of probability
theory.
 When the system enters the
{\em critical} stage, the following relation
holds\cite{NAT01,NAT02,VAR05C}:
\begin{equation}
\Pi ( \omega ) = \frac{18}{5 \omega^2} -\frac{6 \cos \omega}{5
\omega^2} -\frac{12 \sin \omega}{5 \omega^3}. \label{fasma}
\end{equation}
For $\omega \rightarrow 0$, Eq.(\ref{fasma}) leads
to\cite{NAT01,NAT02,newbook}
\[ \Pi (\omega )\approx 1-0.07
\omega^2\] which reflects\cite{VAR05C} that the variance of $\chi$
is given by
\[ \kappa_1=\langle \chi^2 \rangle -\langle \chi \rangle
^2=0.07,\] where $\langle f( \chi) \rangle = \sum_{k=1}^N p_k
f(\chi_k )$.
  The entropy $S$ in the natural time-domain is defined
as\cite{NAT01,NAT03B} \[ S \equiv  \langle \chi \ln \chi \rangle -
\langle \chi \rangle \ln \langle \chi \rangle,\] which  depends on
the sequential order of events\cite{NAT04,NAT05}. It
exhibits\cite{NAT05B} concavity, positivity,
Lesche\cite{LES82,LES04} stability, and for SES activities (critical dynamics)  its value is
smaller\cite{NAT03B,newbook} than the value $S_u (=\ln 2
/2-1/4\approx 0.0966$) of a ``uniform'' (u) distribution (as
defined in Refs. \cite{NAT01,NAT03,NAT03B,NAT04,NAT05}, e.g.
when all $p_k$ are equal or $Q_k$ are positive independent and
identically distributed random variables of finite variance. In
this case, $\kappa_1$ and $S$ are designated $\kappa_u(=1/12)$ and
$S_u$, respectively.). Thus, $S < S_u$. The same holds for the
value of the entropy obtained\cite{NAT05B,NAT06A} upon considering
the time reversal ${\mathcal T}$, i.e., ${\mathcal T}
p_k=p_{N-k+1}$, which is labelled by $S_-$.

In summary, the SES activities, when analyzed in natural time
exhibit {\em infinitely} ranged temporal correlations and  obey
the conditions\cite{NAT06A,NAT06B}:
\begin{equation}\label{eq1}
    \kappa_1 = 0.07
\end{equation}
and
\begin{equation}\label{eq2}
    S, S_- < S_u.
\end{equation}

We first present in subsection \ref{mia} the most recent experimental results
on Seismic Electric Signals activities and their properties are compared,  in
subsection \ref{dia},  with those of the model proposed as well as with a discrete
model proposed\cite{voss} by R.F.Voss(see Appendix B).
The classification, in advance, of these SES activities, on the
basis of relations (\ref{eq1}) and (\ref{eq2}) has been verified
by the occurrence of three magnitude 6.0-class earthquakes (see
Appendix A).

\subsection{The recent electric field data}
\label{mia}

Figure \ref{f3} depicts the original time series of four
electrical disturbances that have been recently recorded on: (a)
February 8, 2007, (b) April 23, 2007, (c) April 24, 2007 and (d)
November 7, 2007 at a measuring station termed Patras (PAT)
located at $\approx 160$km west of Athens. All these four recent
electric signals were analyzed in natural time. For example, if we read in
natural time the signal on April 24, 2007 (Fig.\ref{f3}(c)) -the
dichotomous representation of which is marked by the dotted (blue)
line in Fig.\ref{fn6}(a)- we find the natural time representation
of Fig.\ref{fn6}(b) the analysis of which leads to the values
$\kappa_1= 0.067\pm 0.003$, $S= 0.072 \pm 0.003$, $S_-=0.069\pm
0.003$. The relevant results of all the four signals are compiled
in Table \ref{tab1}
 and
found to be consistent with the conditions (\ref{eq1}) and
(\ref{eq2}), thus they can be classified as SES activities (for
their subsequent seismicity as well as for more recent SES
activities see Appendix A). An inspection of Table \ref{tab1}
shows that the $S$ value is more or less comparable to that of
$S_-$, but experimental uncertainty does not allow any conclusion
which of them is larger. Note that in several former
examples\cite{NAT05B}, the data analysis also showed that the $S$
value may either be smaller or larger than $S_{-}$.

\subsection{Comparison of the SES properties with those of the model proposed}
\label{dia} We now turn to investigate whether the parameters
$\kappa_1$, $S$ and $S_-$ deduced from the 1/f model of Section
\ref{dyo} are consistent to those resulted from the analysis of
the SES activities observed. Figure \ref{natmodel} summarizes the
results of $10^4$ runs of the model which,  for moderate sizes of
$n$, seems to obey  more or less the conditions (\ref{eq1}) and
(\ref{eq2}). In particular, for $n \lesssim 10^2$ (which is
frequently the number of pulses of the SES activities observed in
field experiments), Fig.\ref{natmodel} shows that $\kappa_1$ is
close to 0.070, $S<S_u$ and (in most cases) $S_-<S_u$. A closer
inspection of Fig.\ref{natmodel}, however, reveals the following
incompatibility of the model with the experimental results: For $n
\lesssim 10^2$, the model clearly suggests that $S_-> S$, thus
disagreeing with the experimental data which show, as mentioned
above, that $S$ may either be smaller or larger than $S_-$. The
origin of this incompatibility has not yet been fully understood. It
might be due to the fact that SES activities exhibit {\em
critical} dynamics, while the model cannot capture all the
characteristics of such dynamics.

Since the model proposed here might be considered as reminiscent
of an early $1/f$ model proposed by R.F.Voss (see Ref.\cite{voss}), which is
also a discrete one, we also present in Fig.\ref{voss} the
corresponding results of $10^4$ runs of that model. (The details of the model
are given in Appendix B). A comparison of Figs.\ref{natmodel}
and \ref{voss} for $n\leq 10^2$ reveals that the results of the
two models differ essentially. For example, in the Voss model,
$S_-$ is larger than $S_u$ while the opposite holds for the model
proposed here. Moreover, when comparing the results of the Voss
model with those of the SES activities we also find considerable
differences. For example, the $\kappa_1$ value deduced from the
Voss model is (on the average) {\em larger} than $\kappa_u$, thus
differing considerably from the value $\kappa_1 \approx 0.070$ of
the SES activities. This, as mentioned above, is comparable to the
one deduced from the model proposed here and explains why we
focused on this paper on the comparison of that model -among the
variety of models suggested to date for the explanation of the 1/f
behavior- with the experimental values obtained from the analysis
of the SES activities

\section{Conclusions}
\label{tesera} In summary, using the newly introduced concept of
natural time:(a) A simple model is proposed that exhibits $1/f^a$
behavior with $a$ close to unity.  (b) Electric signals, recorded
during the last few months in Greece, are classified as SES
activities since they exhibit {\em infinitely} ranged temporal
correlations. Actually, three magnitude 6.0 class earthquakes
already occurred in Greece (see Appendix A). (c) For sizes $n$
comparable to those of the SES activities measured in the field
experiments (i.e., $n\lesssim 10^2$), the model proposed here
leads to values of the parameters $\kappa_1$ ($\approx 0.070$) and
$S, S_-$ ($< S_u$) that are consistent with those deduced from the
SES activities analysis. Despite of this fact, however, the model
results in  $S_-$ values that are almost always larger than those
of $S$, while the observed SES activities result in $S$ values
that may be either larger or smaller than $S_-$. This discrepancy
might be due to the inability of the model to capture the
characteristics of {\em critical} dynamics which is exhibited by
SES activities.

\appendix

\section{What happened after the SES activities depicted in Fig.\ref{f3}}

We clarify that, during the last decade, preseismic
information\cite{vers4}  based on SES activities is issued {\em
only} when the magnitude of the strongest EQ of the impending EQ
activity is estimated -by means of the SES
amplitude\cite{proto,var86b,var88x,VAR91,VAR93}- to be  comparable
to 6.0 units or larger\cite{newbook}. Here, in the first two
subsections, we explain what happened after the SES activity at
PAT on February 8, 2007 (see Fig.\ref{f3}(a)) and on April 23 and
24, 2007 (see Figs.\ref{f3}(b),(c)). The relevant analysis of
seismicity after the SES activity on November 7, 2007, which was
still in progress\cite{var07x} during the initial submission of
this paper on 23 November, 2007 and hence completed  afterwards,
is presented in Subsection 3.

\subsection{What happened after the SES activity of February 8, 2007}

According to the National Observatory of Athens, NOA (the seismic
data of which will be used here),  a strong earthquake (EQ) with
magnitude 6.0-units  occurred at Kefallonia area, i.e., $38.34^o$N
$20.42^o$E, at 13:57 UT on March 25, 2007. We show below that the
occurrence time of this strong EQ
 can be estimated by following the procedure described in
Refs.\cite{NAT01,VAR05C,NAT06A,NAT06B,newbook}.

We study how the seismicity evolved after the recording of the SES
activity at PAT on February 8, 2007, by considering either the
area A:$N_{37.6}^{39.0}E_{20.0}^{22.2}$ or its smaller area
B:$N_{37.6}^{38.6}E_{20.0}^{22.2}$ (see Fig.\ref{af1}). If we set
the natural time for seismicity zero at the initiation of the  SES
activity on February 8, 2007, we form time series of seismic
events in natural time for various time windows as the number $N$
of consecutive (small) EQs increases. We then compute the
normalized power spectrum in natural time $\Pi (\phi )$ for each
of the time windows. Excerpt of these results, which refers to the
values deduced during the period from 20:53:59 UT on March 19,
2007, to 11:56:30 UT on 25 March, 2007, is depicted in red in
Fig.\ref{af2}(a).  This figure corresponds to the area B  with
magnitude threshold (hereafter referring to the local magnitude ML
or the `duration' magnitude MD) $M_{thres}=3.2$. In the same
figure, we plot in blue the power spectrum obeying the relation
(2) which holds, as mentioned, when the system enters the {\em
critical} stage. The date and the time of the occurrence of each
small earthquake (with magnitude exceeding (or equal to) the
aforementioned threshold) that occurred in area  B, is also
written in red in each panel. An inspection of this figure reveals
that the red crosses  approach the blue line as $N$ increases and a
{\em coincidence} occurs at the last small event which had a
magnitude 3.2 and occurred at 11:56:30 UT on March 25, 2007, i.e.,
just two hours before the strong 6.0 EQ. To ensure that this
coincidence is a {\em true} one (see also below) we also calculate
the evolution of the quantities $\kappa_1$,$S$ and $S_{-}$   and
the results are depicted in Fig. \ref{af2}(b) and \ref{af2}(c) for
the same magnitude thresholds  for each of the areas B and A,
respectively.

The conditions for a coincidence to be considered as {\em true}
are the following (e.g., see
Refs.\cite{NAT01,VAR05C,NAT06A,NAT06B,newbook}): First, the
`average' distance $\langle D \rangle$ between the empirical and
the theoretical $\Pi(\phi )$(i.e., the red crosses and the blue line,
respectively, in Fig.\ref{af2}(a)) should be smaller than
$10^{-2}$. See Fig. \ref{af2}(b),(c) where we plot $\langle D
\rangle$ versus the conventional time for the aforementioned two
areas B and A, respectively. Second, in the examples observed to
date, a few events {\em before} the coincidence leading to the
strong EQ, the evolving $\Pi(\phi )$ has been found to approach
that of the relation (2), i.e., the blue one in Fig.\ref{af2}(a) ,
from {\em below} (cf. this reflects that during this approach the
$\kappa_1$-value decreases as the number of events increases). In
addition, both values $S$ and $S_{-}$ should be smaller than $S_u$
at the coincidence. Finally, since the process concerned is
self-similar ({\em critical} dynamics), the time of the occurrence
of the (true) coincidence should {\em not} change, in principle,
upon changing the surrounding area (and the magnitude threshold
$M_{thres}$) used in the calculation. Note that in Fig.
\ref{af2}(b), upon the occurrence of the aforementioned last small
event at 11:56:30 UT of March 25, 2007, in area B  the $\langle D
\rangle$ value becomes smaller than $10^{-2}$. The same was found
to hold for the area A, see Fig.\ref{af2}(c).

\subsection{What happened after the SES activities of April 23 and 24, 2007}

We investigate the seismicity after the aforementioned SES
activities depicted in Figs.\ref{f3}(b) and \ref{f3}(c). The
investigation is made in the areas C:
$N^{39.4}_{37.6}E^{22.2}_{20.2}$ and  D:
$N^{39.4}_{38.1}E^{22.2}_{20.2}$ (see Fig.\ref{afx1}).  Starting
the computation of seismicity from the initiation of the SES
activity on April 24, 2007 (which, between the two SES activities
depicted in Figs.\ref{f3}(b) and \ref{f3}(c), has the  higher
actual amplitude), we obtain the results depicted in
Figs.\ref{afx2}(a),(b) and \ref{afx2}(c),(d) for the areas C and
D, respectively, for $M_{thres}=3.0$ and $M_{thres}=3.2$.   An
inspection of the parameters $\langle D \rangle$, $\kappa_1$,$S$
and $S_-$ reveals that they exhibited a {\em true} coincidence (as
discussed above) around  June 25, 2007, i.e., around four days
before the 5.8 EQ that occurred at 18:09:11 UT on June 29, 2007,
with an epicenter at $39.3^o$N $20.3^o$E (shown by a star in
Fig.\ref{afx1}).

\subsection{Study of the seismicity after the SES activity on November 7, 2007}
This study, which as mentioned above was still in progress upon
the initial submission of this paper, was made by investigating
the seismicity in the area B of Fig.\ref{af1}, i.e.,
$N^{38.6}_{37.6}E^{22.2}_{20.0}$ as well as in the larger area,
i.e., $N^{38.6}_{37.6}E^{23.3}_{20.0}$ (see Ref.\cite{var07x}),
which are shown in Fig.\ref{fig12}. The parameters $\kappa_1$,
$S$, $S_-$ and $\langle D \rangle$ computed during the subsequent
period for $M_{thres}=3.1$ are depicted in Figs \ref{fig13}(a) and
\ref{fig13}(b) for the smaller and larger area, respectively. An
inspection of these two figures reveals that the conditions for a
{\em true} coincidence (see subsection A.1) were obeyed upon the
occurrence of the first event early in the morning (i.e., at 03:25
UT) on 30 December, 2007 with an epicenter at $37.8^o$N $20.2^o$E
and magnitude 3.9. Almost three hours later, i.e., at 06:42 UT, a
strong earthquake of magnitude 5.3 occurred at $37.6^o$N $20.9^o$E
marked with the small star in Fig.\ref{fig14}. In addition, and
quite interestingly, the four parameters $\kappa_1$, $S$, $S_-$
and $\langle D \rangle$ during the next few days continued to
fulfill  the conditions for a true coincidence, as it is evident
from a closer inspection of Figs.\ref{fig13}(a) and
\ref{fig13}(b). Actually, at 05:14 UT on 6 January, 2008, a major
magnitude 6.6 earthquake occurred, which was felt not only all
over Greece but also in adjacent countries, e.g., southern Italy
and western Turkey. Its epicenter, marked with a large star in
Fig.\ref{fig14}, was located at $37.1^o$N $22.8^o$E, i.e., only
around 50 km to the south of the larger area studied since the
initial submission of this paper. Interestingly, in
the part of the latter area
 adjacent to the epicenter (which is shaded in Fig.\ref{fig14})
the four parameters $\kappa_1$, $S$, $S_-$ and $\langle D \rangle$
reached the conditions for a true coincidence just in the morning
(i.e., at 04:24 UT) of 5 January, 2008  upon the occurrence of a
magnitude 3.6 earthquake at $38.4^o$N $22.0^o$E (it corresponds to
the last point in Fig.\ref{fig13}(c) where $\langle D \rangle$
becomes smaller than $1\%$).

Finally, we note that {\em after} the initial submission of this
paper, two additional SES activities have been recorded as follows
(see Fig.\ref{NFX}): One SES activity at PAT on 10 January, 2008
and another one on 14 January, 2008 at the station PIR located in
western Greece, see Fig.\ref{fig12} (cf. The configuration of the
measuring dipoles in the latter station is described in detail in
the EPAPS document of Ref.\cite{NAT06B}). Their subsequent
seismicities are currently studied along the lines explained above
considering the evolving seismicity in the following areas:
Concerning the former SES activity at PAT the areas depicted in
Fig.\ref{fig12}, while for the one at PIR on 14 January, 2008, the
subsequent seismicity is studied in the area B of Fig.\ref{af1} as
well as in the larger area $N^{38.6}_{36.0}E^{22.5}_{20.0}$ and in
the one surrounding the epicenter\cite{acta06} ($36^o$N $23^o$E).

We now offer some comments on the  classification of the
aforementioned electric signals of Fig.\ref{NFX} as SES
activities. Concerning the signal on 14 January, 2008, which is of
clear dichotomous nature, the analysis is made by considering that
$Q_k$ stands for the duration of $k$-th pulse, as mentioned in
Section I, and the following parameters are obtained:
$\kappa_1=0.070\pm 0.005$, $S=0.086\pm 0.003$, $S_-=0.070\pm 0.005$, which obey the
conditions (\ref{eq1}) and (\ref{eq2}) for the classification of
the signal as SES activity. Furthermore, note the $S_-$ is smaller
than that of $S$, which is not compatible with the model proposed
in Section II, thus strengthening the point mentioned in Section
III.B as well as in the Conclusions (Section IV) that the model
does not seem to capture the characteristics of critical dynamics
exhibited by SES activities. We now turn to the signal recorded at
PAT on 10 January, 2008, the feature of which is not clearly
dichotomous since it consists of three main pulses that seem to
overlap (Note that, in general, if pulses of very short duration
exist, the calculation of $Q_k$ should necessarily consider the
characteristics of the low pass filters used in our measurements\cite{PRL03,newbook}.
This has been considered in drawing the
dichotomous representation -marked by the dotted (blue) line- in
Fig.\ref{NFX}(a)). Its analysis leads to the following values
$\kappa_1=0.070\pm 0.010$, $S=0.050\pm 0.010$, $S_-=0.060\pm 0.010$. These values, which are
{\em different} from those deduced from the analysis of the SES
activity on 7 November, 2007, also obey the conditions (\ref{eq1})
and (\ref{eq2}). At this point, we clarify that the optimality of
natural time domain for enhancing the signal's localization in the
time-frequency space was shown\cite{ABE05} without assuming that
$Q_k$ stands for the pulse duration, but it was noted that in
general $Q_k$ is a quantity proportional to the corresponding
energy released in the $k$-th event (estimated by means of the
time integration of the signal exceeding the level of the
irrelevant background noise).

\subsection{Additional data}
\begin{figure*}
\includegraphics{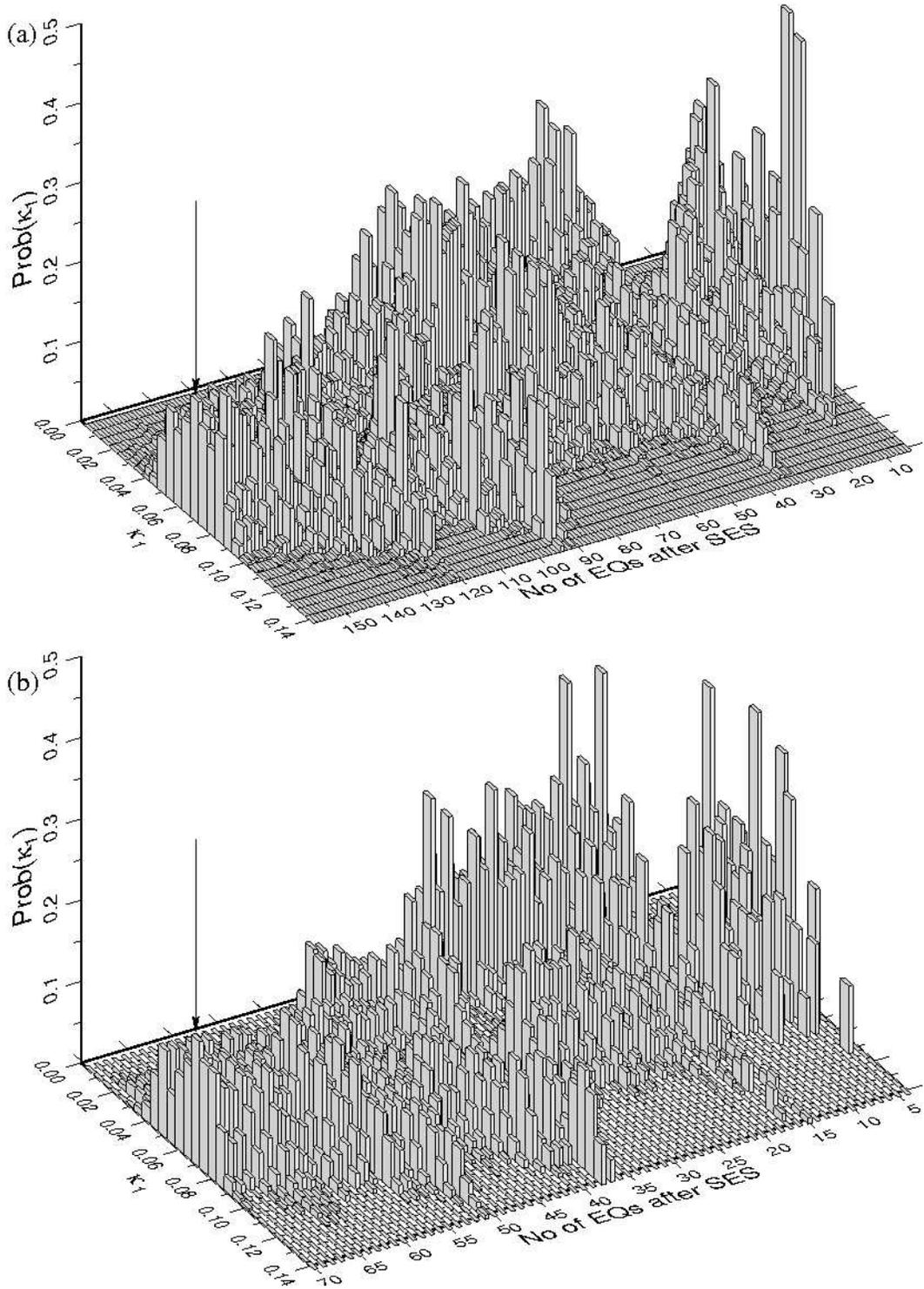}
\caption{(a) Prob($\kappa_1$) versus $\kappa_1$ of the seismicity,
for $M_{thres}=3.0$, subsequent to the SES activity at PAT on
October 9, 2008. The last event corresponds to the $M_L=3.0$ event
that occurred at 00:43 UT on December 6, 2008 at $37.81^oN
21.15^oE$. (b)The same as in (a), but for
$M_{thres}=3.2$}\label{fig18}
\end{figure*}

Meanwhile, the earthquakes that followed the SES activities: (i)
depicted in Fig.\ref{NFX} and (ii) recorded on 29 February - 2
March, 2008 later at PIR (see N.V.Sarlis, E.S. Skordas, M.S.
Lazaridou, arxiv:0802.3329v4), have been described in N.V.Sarlis,
E.S. Skordas, M.S. Lazaridou, Proc. Jpn. Acad. B 84, 331, 2008. In
addition, a strong electrical disturbance has been recorded at 9
October, 2008 (Fig.\ref{R091008}), having an amplitude comparable
to that in Fig.\ref{f3}(d). The relevant analysis of the
subsequent seismicity in natural time is carried out in the area
$N_{37.5}^{38.6} E_{19.8}^{23.3}$. Such an analysis, when
considering the NOA seismic data published until December 6, 2008,
we find the results depicted in Figs. \ref{fig18}(a) and
\ref{fig18}(b) for magnitude thresholds 3.0 and 3.2, respectively.
An inspection of these figures shows that Prob( $\kappa_1$)
exhibits a local maximum at $\kappa_1 \approx 0.070$, thus obeying
Eq.\ref{eq1}. The spatial invariance of this result is currently
investigated in order to check whether the critical point has been
actually approached.

%\begin{figure}
%\includegraphics{FigX3.2.eps}
%\caption{The same as Fig.\ref{fig18} but for
%$M_{thres}=3.2$}\label{fig19}
%\end{figure}

\section{Comparison of the present model with a model suggested by R.F.Voss}
A model producing $1/f$ noise, suggested by Richard F. Voss, was
presented  in Ref.\cite{voss}. This model assumes some number $m$
of $k$-sided dices, i.e., the outcome of a rolling the $l$-th dice
is equally distributed among the values $d_l=1,2,\ldots k$. To
each dice $l$, one relates the $l$-th digit of the binary
representation of a number $n=0,1,\ldots ,2^{m}-1$. The procedure
to generate the $1/f$ noise starts ($n=0$) by rolling all dices and
assign their sum to $v_0$($=\sum_{l=1}^m d_l$). In the second step
($n=1$), one rerolls only the dice associated with the least significant digit
of the binary representation, and the new roll $d'_1$ is summed
with the previous rolls of all other dices so that
$v_1=d'_1+\sum_{l=2}^m d_l$. The procedure continues up to
$n=2^{m}-1$, each time rolling only the dices associated with the
digits that change when considering the binary representation of
$n$  compared to $n-1$. This model results in time-series $v_n$ of
length $2^m$ whose spectrum is close to 1/f, and their values
$v_n$ are obviously distributed among the integers $m$ and $mk$.
The time series $v_n$ clearly differ from $\epsilon_n$ apart from
the fact that they both have integer values. To visualize this
difference, we depict in Fig.\ref{pdf}(b) the distribution of
$p(v_{2^{m-1}})$ for the case of  6-sided dices ($k=6$). It can be
easily found,  since upon considering the binary representation of
$n=2^{m-1}$ compared to $n-1=2^{m-1}-1$,  {\em all}  digits
change and one rerolls all dices. Clearly, rolling all dices
results in a symmetric ({\em skewnessless}) distribution for
$v_{2^{m-1}}$ (see Fig.\ref{pdf}(b)). Indeed, by considering the
distribution of the sum of rolling $m$ independent $k$-sided dices
and using the characteristic function method discussed in
subsection \ref{dyoB}, one can find the following cumulants:
\begin{eqnarray}
\langle v_{2^{m-1}} \rangle  =  m\frac{k+1}{2}, \label{nc1} \\
\langle \left( v_{2^{m-1}}  -\langle v_{2^{m-1}} \rangle
\right)^2 \rangle  =  m \frac{(k^2-1)}{12}, \label{nc2} \\
\langle \left( v_{2^{m-1}}  -  \langle v_{2^{m-1}} \rangle
\right)^3\rangle=0, \label{nc3} \\
\langle \left( v_{2^{m-1}}  -\langle v_{2^{m-1}} \rangle \right)^4
\rangle-3\langle \left( v_{2^{m-1}}  -\langle v_{2^{m-1}} \rangle
\right)^2 \rangle^2=m\frac{(1-k^4)}{120}. \label{nc4}
\end{eqnarray}
Clearly, Eqs.(\ref{nc1}) to (\ref{nc4}) for the distribution of
$v_{2^{m-1}}$ differ from Eqs.(\ref{t1}) to (\ref{t4}) for the
distribution of $\epsilon_n$. Among their differences,  the
following two are the most striking: First,  $v_{2^{m-1}}$ is
skewnessless (see Eq.(\ref{nc3})) whereas that of  $\epsilon_n$ is not, and second  the
two distributions have different signs in their kurtoses.

%\section*{References}
%\bibliography{newrefa}
\bibliographystyle{apsrev}

\end{document}